\documentclass[twocolumn]{jpsj3}
\usepackage{txfonts}

\title{
Gutzwiller Method for an Extended Periodic Anderson Model\\
with the $c$-$f$ Coulomb Interaction  
}

\author{Katsunori \surname{Kubo}}
\inst{
Advanced Science Research Center, Japan Atomic Energy Agency,
Tokai, Ibaraki 319-1195}

\recdate{August 11, 2011;
accepted August 25, 2011;
published online October 17, 2011}

\abst{
We study an extended periodic Anderson model
with the Coulomb interaction $U_{cf}$
between conduction and $f$ electrons
by the Gutzwiller method.
The crossovers between the Kondo, intermediate-valence,
and almost empty $f$-electron regimes become sharper with $U_{cf}$,
and for a sufficiently large $U_{cf}$,
become first-order phase transitions.
In the Kondo regime, a large enhancement in the effective mass
occurs as in the ordinary periodic Anderson model without $U_{cf}$.
In addition, we find that a large mass enhancement
also occurs in the intermediate-valence regime
by the effect of $U_{cf}$.
}

\kword{mass enhancement, Gutzwiller approximation,
  extended periodic Anderson model, valence transition,
  valence fluctuations, heavy-fermion superconductivity
}

\begin{document}
\maketitle

\section{Introduction}

In rare-earth and actinide compounds,
several interesting phenomena,
such as
magnetism, heavy-fermion phenomena, and superconductivity,
occur owing to the interplay of
the strong Coulomb interaction $U$ between $f$ electrons
and the hybridization $V$ between the localized $f$-orbital
and conduction band.

Among such phenomena,
heavy-fermion superconductivity
has been one of the central issues in $f$-electron physics
after the discovery of the superconductivity in
CeCu$_2$Si$_2$.~\cite{Steglich1979}
In the heavy-fermion systems,
the conventional, phonon-mediated, $s$-wave superconductivity
is hardly realized owing to the strong onsite Coulomb interaction.
Then, pairing mechanisms other than the phonon-mediated mechanism
have been discussed.
The magnetic-fluctuation-mediated superconducting mechanism
may be common in heavy-fermion superconductors,
since superconductivity is realized near a magnetic quantum critical point
in many compounds.

However, some heavy-fermion superconductors
are difficult to understand solely by the magnetic fluctuation scenario.
For example,
the superconducting transition temperatures under pressure
in CeCu$_2$Si$_2$~\cite{Bellarbi} and CeCu$_2$Ge$_2$,\cite{Vargoz1998}
become maximum far away from the magnetic quantum critical points.
In CeCu$_2$Si$_{1.8}$Ge$_{0.2}$,~\cite{Yuan2003}
the superconducting region splits into two regions:
the low-pressure region close to the magnetic critical point
and
the high-pressure region away from the magnetic critical point.
To explain the high-pressure superconducting phase,
a valence fluctuation scenario is
proposed~\cite{Miyake1999,Onishi2000JPSJ,Watanabe2006,Sugibayashi2008}
and the importance of the Coulomb interaction $U_{cf}$
between conduction and $f$ electrons has been discussed
in addition to $U$ and $V$.

Valence fluctuations
are expected from the rapid change in the valence of an $f$ ion
in these compounds under pressure,
which is suggested from the behavior of the effective mass $m^*$.
In the periodic Anderson model (PAM),
$m^*$ and the number of electrons, $n_f$, in the $f$ orbital per site
follow the relation~\cite{Rice1986,Fazekas1987}
\begin{equation}
  \frac{m^*}{m}=\frac{1-n_f/2}{1-n_f},
  \label{eq:Rice-Ueda}
\end{equation}
where $m$ is the free-electron mass.
This relation is derived using the Gutzwiller method
for $U \rightarrow \infty$.
Thus, a large change in $m^*$
indicates a large change in $n_f$.
For CeCu$_2$Si$_2$ and CeCu$_2$Ge$_2$,
$m^*$ is experimentally deduced from specific heat measurements
or the temperature dependence of electrical resistivity,
and it is found that $m^*$ decreases rapidly at approximately the pressure
where the superconducting transition temperature
becomes maximum.~\cite{Jaccard1999,Holmes2004}
Thus, these observations indicate that a sharp valence change or
large valence fluctuations play important roles in superconductivity.

However, there are problems with the relation between
the effective mass and valence.
First, eq.~\eqref{eq:Rice-Ueda} is derived
for an ordinary PAM,
which does not show a sharp valence change.
Thus, we cannot naively apply eq.~\eqref{eq:Rice-Ueda}
to a system with large valence fluctuations.
Second,
the effective mass has a peak in CeCu$_2$Si$_2$
under pressure
before the superconducting transition temperature becomes maximum,
that is,
the effective mass varies nonmonotonically under pressure.~\cite{Holmes2004}
%
This nonmonotonic variation in the effective mass
cannot be explained by eq.~\eqref{eq:Rice-Ueda},
since the $n_f$ of a Ce ion is expected to decrease monotonically under pressure
for the following reasons.
A positively charged Ce ion should be surrounded
by negative charges.
When these negative charges get close to a Ce ion
under pressure, the $f$ level $\epsilon_f$ of Ce is lifted.
Under pressure, the overlap between the wave functions
of the $f$ orbital and the conduction band
increases, and $V$ increases.
Both the effects of pressure on $\epsilon_f$ and $V$
result in a decrease in $n_f$.
Indeed, a monotonic decrease in $n_f$ under pressure has been observed
in CeCu$_2$Si$_2$ by an X-ray absorption experiment recently.~\cite{Rueff2011}
We also note that, in CeCu$_2$Ge$_2$,
the pressure dependence of the effective mass has a shoulder structure
before the superconducting transition temperature
becomes maximum.~\cite{Jaccard1999}
This shoulder structure may also become a peak as in CeCu$_2$Si$_2$,
if we subtract the contributions of magnetic fluctuations,
which are large in the low-pressure region.

The peak structures in the effective mass may be explained by
considering a combined effect
of the renormalization described by eq.~\eqref{eq:Rice-Ueda}
and valence fluctuations.~\cite{Holmes2004}
However, the applicability of eq.~\eqref{eq:Rice-Ueda}
to a system with large valence fluctuations is not justified.
Thus, it is an interesting problem
how eq.~\eqref{eq:Rice-Ueda} can be extended
to a model that shows a sharp valence change.
A study of such a problem will be helpful
to understand superconductivity in
CeCu$_2$Si$_2$ and CeCu$_2$Ge$_2$
by the valence fluctuation scenario.

Note that heavy-fermion behaviors in
$\alpha$-YbAlB$_4$ and $\beta$-YbAlB$_4$~\cite{Macaluso2007}
are also difficult to explain by eq.~\eqref{eq:Rice-Ueda}.
The valences of the Yb ion are $+2.73$ for $\alpha$-YbAlB$_4$
and $+2.75$ for $\beta$-YbAlB$_4$,~\cite{Okawa2010}
that is, the hole numbers of Yb ions are
$n_f=0.73$ and 0.75 for $\alpha$-YbAlB$_4$ and $\beta$-YbAlB$_4$, respectively.
For these values of $n_f$, i.e., much less than unity,
we cannot expect heavy-fermion behavior from eq.~\eqref{eq:Rice-Ueda}.

In this research,
we study an extended periodic Anderson model (EPAM) with the Coulomb interaction
$U_{cf}$ between conduction and $f$ electrons by the Gutzwiller method.
We extend the Gutzwiller method for the PAM
developed by Fazekas and Brandow~\cite{Fazekas1987} to the present model.
Then, we investigate the effect of $U_{cf}$ on the effective mass.
Although the EPAM
has been studied as a typical model
for valence transition~\cite{Goncalves1975}
and
has been investigated by some modern techniques
in recent years~\cite{Onishi2000JPSJ,Onishi2000PhysicaB,Watanabe2006,Sugibayashi2008,Saiga2008,Yoshida2011}
after the proposal of the valence fluctuation scenario for superconductivity,
the effect of $U_{cf}$ on the mass enhancement has not been clarified well.
Some of the results have already been reported in our previous paper;~\cite{Kubo2011}
here, we report the details of the method
and also add new results.

\section{Formulation}

The EPAM is given by~\cite{Goncalves1975}
\begin{equation}
  \begin{split}
    H=&\sum_{\mib{k} \sigma}\epsilon_{\mib{k}}
    c^{\dagger}_{\mib{k} \sigma}c_{\mib{k} \sigma}
    +\epsilon_f \sum_{\mib{r} \sigma}n_{f \mib{r} \sigma}\\
    &-V\sum_{\mib{k} \sigma}(f^{\dagger}_{\mib{k} \sigma}c_{\mib{k} \sigma}
                            +c^{\dagger}_{\mib{k} \sigma}f_{\mib{k} \sigma})\\
    &+U\sum_{\mib{r}}n_{f \mib{r} \uparrow}n_{f \mib{r} \downarrow}
    +U_{cf}\sum_{\mib{r} \sigma \sigma^{\prime}}
    n_{c \mib{r} \sigma}n_{f \mib{r} \sigma^{\prime}},
  \end{split}
\end{equation}
where $c_{\mib{k} \sigma}$ and $f_{\mib{k} \sigma}$ are
the annihilation operators
of conduction and $f$ electrons, respectively,
with the momentum $\mib{k}$ and the spin $\sigma$.
$n_{c \mib{r} \sigma}$ and $n_{f \mib{r} \sigma}$ are the number operators
at site $\mib{r}$ with $\sigma$ of the conduction and $f$ electrons,
respectively.
$\epsilon_{\mib{k}}$ is the kinetic energy of the conduction electron.
We have not taken orbital degrees of freedom into consideration.
This simplification may be justified
for a system with tetragonal symmetry
such as CeCu$_2$Si$_2$ and CeCu$_2$Ge$_2$
and
for a system with orthorhombic symmetry
such as $\alpha$-YbAlB$_4$ and $\beta$-YbAlB$_4$,
since the crystalline electric field ground states of $f$ electrons
are Kramers doublets in these systems.
In the following, we set the energy level of the conduction band
as the origin of energy, i.e., $\sum_{\mib{k}}\epsilon_{\mib{k}}=0$.
We set $U \rightarrow \infty$, since the onsite Coulomb interaction
between well-localized $f$ electrons is large.

We consider the variational wave function
given by
\begin{equation}
  | \psi \rangle=P_{ff}P_{cf} | \phi \rangle,
  \label{eq:psi}
\end{equation}
where
\begin{equation}
  P_{ff}=\prod_{\mib{r}}[1-n_{f \mib{r} \uparrow}n_{f \mib{r} \downarrow}]
\end{equation}
excludes the double occupancy of $f$ electrons at the same site,
and
\begin{equation}
  P_{cf}=\prod_{\mib{r} \sigma \sigma^{\prime}}
  [1-(1-g)n_{c \mib{r} \sigma}n_{f \mib{r} \sigma^{\prime}}]
\end{equation}
is introduced to deal with the onsite correlation between conduction
and $f$ electrons.~\cite{Onishi2000PhysicaB}
$g$ is a variational parameter.
The one-electron part of the wave function is given by
\begin{equation}
  | \phi \rangle=
  \prod_{k<k_{\text{F}}, \sigma}
  [c^{\dagger}_{\mib{k} \sigma}+a(\mib{k})f^{\dagger}_{\mib{k} \sigma}]
  | 0\rangle,
\end{equation}
where $k_{\text{F}}$ is the Fermi momentum
for the free conduction band without $f$ electrons,
$|0\rangle$ denotes vacuum,
and $a(\mib{k})$ is determined variationally.
Here, we have assumed that the total number $n$ of electrons per site
is less than 2.

In the method by Fazekas and Brandow,~\cite{Fazekas1987}
the $f$-electron state is expanded in the basis state in real space,
since it is convenient to deal with the projection operator $P_{ff}$.
In the present study, we consider $P_{cf}$ in addition to $P_{ff}$,
and thus we also expand the conduction-electron state
in real space.
This is the main difference from the method of Fazekas and Brandow.
The creation operator is expanded as
\begin{equation}
  b^{\dagger}_{\mib{k}}
  =\frac{1}{\sqrt{L}}\sum_{\mib{r}}
  e^{i\mib{k}\cdot\mib{r}}b^{\dagger}_{\mib{r}}
  =\sum_{\mib{r}}\varphi_{\mib{k}}(\mib{r})b^{\dagger}_{\mib{r}},
\end{equation}
where $L$ is the number of lattice sites
and $b$ denotes $c_{\sigma}$ or $f_{\sigma}$.
Then, the basis state in momentum space is expanded as
\begin{equation}
  \begin{split}
    |\{ \mib{k}^{(b)} \} \rangle
    &=\prod^{N_b}_{i=1} b^{\dagger}_{\mib{k}_i} |0\rangle\\
    &=\sum_{\{ \mib{r}^{(b)} \}}\det[\varphi_{\mib{k}^{(b)}}(\mib{r}^{(b)})]
    \prod^{N_b}_{i=1} b^{\dagger}_{\mib{r}_i} |0\rangle\\
    &=\sum_{\{ \mib{r}^{(b)} \}}\det[\varphi_{\mib{k}^{(b)}}(\mib{r}^{(b)})]
    |\{ \mib{r}^{(b)} \} \rangle.
  \end{split}
  \label{eq:basis_state}
\end{equation}
The determinant is defined as
\begin{equation}
  \det[\varphi_{\mib{k}}(\mib{r})]
  =
  \begin{vmatrix}
    \varphi_{\mib{k}_1}(\mib{r}_1) & \varphi_{\mib{k}_1}(\mib{r}_2) & \\
    \varphi_{\mib{k}_2}(\mib{r}_1) & \varphi_{\mib{k}_2}(\mib{r}_2) & \\
    & & \ddots
  \end{vmatrix}.
\end{equation}
The basis state including both $c$ and $f$ electrons is given by
\begin{equation}
  | \{\mib{r}^{(c)}\} \{\mib{r}^{(f)}\} \rangle
  =\prod_{\sigma}
  \prod^{N_{c \sigma}}_{j=1} c^{\dagger}_{\mib{r}^{(c \sigma)}_{j}}
  \prod^{N_{f \sigma}}_{i=1} f^{\dagger}_{\mib{r}^{(f \sigma)}_{i}}
  | 0 \rangle,
\end{equation}
where $N_{c \sigma}$ and $N_{f \sigma}$ are
the numbers of conduction and $f$ electrons, respectively,
with spin $\sigma$.
The total number of spin-$\sigma$ electrons,
$N_{\sigma}=N_{c \sigma}+N_{f \sigma}$, should be fixed.
Here, we have introduced the notation
$\{\mib{r}^{(c)}\}=\{\mib{r}^{(c \uparrow)}, \mib{r}^{(c \downarrow)}\}$
and
$\{\mib{r}^{(f)}\}=\{\mib{r}^{(f \uparrow)}, \mib{r}^{(f \downarrow)}\}$.
Then, the one-particle part is expanded as
\begin{equation}
  \begin{split}
    | \phi \rangle
    =&\prod_{\sigma}
    \sum_{N_{f \sigma}, \{\mib{k}^{(f \sigma)}\}}
    (-1)^{\text{perm}(\{\mib{k}^{(f\sigma)}\})} \\
    &\times
    \prod^{N_{c \sigma}}_{j=1} c^{\dagger}_{\mib{k}^{(c \sigma)}_j}
    \prod^{N_{f \sigma}}_{i=1}
    a(\mib{k}^{(f \sigma)}_i) f^{\dagger}_{\mib{k}^{(f \sigma)}_i}
    | 0 \rangle \\
    =&\prod_{\sigma}
    \sum_{N_{f \sigma}, \{\mib{k}^{(f \sigma)}\}}
    (-1)^{\text{perm}(\{\mib{k}^{(f\sigma)}\})} \\
    &\times
    \prod^{N_{f \sigma}}_{i=1} a(\mib{k}^{(f \sigma)}_i) \\
    &\times
    \sum_{\{\mib{r}^{(c \sigma)}\}, \{\mib{r}^{(f \sigma)}\}}
    \det[\varphi_{\mib{k}^{(c \sigma)}}(\mib{r}^{(c \sigma)})]
    \det[\varphi_{\mib{k}^{(f \sigma)}}(\mib{r}^{(f \sigma)})] \\
    &\times
    | \{\mib{r}^{(c)}\} \{\mib{r}^{(f)}\} \rangle,
  \end{split}
  \label{eq:phi}
\end{equation}
where $(-1)^{\text{perm}(\{\mib{k}^{(f\sigma)}\})}$ is the sign
due to the fermion anticommutation relation.
In the summation, we should keep
$k^{(c \sigma)}_j, k^{(f \sigma)}_i<k_{\text{F}}$
and $\{\mib{k}^{(c \sigma)}\} \cap \{\mib{k}^{(f \sigma)}\} = \emptyset$.
Then the projection in eq.~\eqref{eq:psi}
is carried out by restricting the summation in eq.~\eqref{eq:phi}
with the condition
$\{\mib{r}^{(f \uparrow)}\} \cap \{\mib{r}^{(f \downarrow)}\} = \emptyset$
and by multiplying each term by $g^D$.
$D=D_{c \uparrow}+D_{c \downarrow}$,
where $D_{c \sigma}$ is the number in the set
$\{\mib{r}^{(c \sigma)}\} \cap \{\mib{r}^{(f)}\}$.
That is, $D$ is the number of interacting electron pairs through $U_{cf}$.
In the following formulation,
we impose these restrictions without mentioning them explicitly.

Then, we evaluate the normalization factor $\langle \psi | \psi \rangle$
using the approximation introduced in Appendix~\ref{sec:determ}.
By using eq.~\eqref{eq:orthogonal}, we obtain
\begin{equation}
  \begin{split}
    \langle \psi | \psi \rangle
    \simeq&
    \sum_{N_{f \uparrow} N_{f \downarrow}}
    \sum_{\{ \mib{k}^{(f)} \}}
    \prod^{N_{f \uparrow}}_{i=1} a^2(\mib{k}^{(f\uparrow)}_i)
    \prod^{N_{f \downarrow}}_{j=1} a^2(\mib{k}^{(f\downarrow)}_j)\\
    &\times \sum_{\{ \mib{r} \}}
    g^{2D}
    \lvert\det[\varphi_{\mib{k}^{(c \uparrow)}}(\mib{r}^{(c \uparrow)})]
    \rvert^2
    \lvert\det[\varphi_{\mib{k}^{(f \uparrow)}}(\mib{r}^{(f \uparrow)})]
    \rvert^2\\
    &\times
    \lvert\det[\varphi_{\mib{k}^{(c \downarrow)}}(\mib{r}^{(c \downarrow)})]
    \rvert^2
    \lvert\det[\varphi_{\mib{k}^{(f \downarrow)}}(\mib{r}^{(f \downarrow)})]
    \rvert^2,
  \end{split}
\end{equation}
where
$\{\mib{k}^{(f)}\} = \{\mib{k}^{(f\uparrow)}, \mib{k}^{(f\downarrow)}\}$
and
$\{\mib{r}\} = \{\mib{r}^{(c)}, \mib{r}^{(f)}\}$.
By applying eq.~\eqref{eq:average},
we further approximate $\langle \psi | \psi \rangle$ and obtain
\begin{equation}
  \begin{split}
    \langle \psi | \psi \rangle
    \simeq
    \sum_{N_{f \uparrow} N_{f \downarrow} D_{c \uparrow} D_{c \downarrow}}
    &g^{2D}
    X(N_{f \uparrow},N_{f \downarrow})\\
    \times&
    Y(N_f,N_{c \uparrow},D_{c \uparrow})
    Y(N_f,N_{c \downarrow},D_{c \downarrow})\\
    \times&
    Z(N_{f \uparrow})Z(N_{f \downarrow}),
  \end{split}
\end{equation}
where $N_f=N_{f \uparrow}+N_{f \downarrow}$,
\begin{equation}
  X(N_{f \uparrow},N_{f \downarrow})
  =\frac{_{L-N_{f \uparrow}}C_{N_{f \downarrow}}}{_{L}C_{N_{f \downarrow}}},
\end{equation}
\begin{equation}
  Y(N_f,N_{c \sigma},D_{c \sigma})
  =\frac{_{N_f}C_{D_{c \sigma}} {}_{L-N_f}C_{N_{c \sigma}-D_{c \sigma}}}
  {_LC_{N_{c \sigma}}},
\end{equation}
and
\begin{equation}
  \begin{split}
    Z(N_{f \sigma})
    &=\sum_{\{ \mib{k}^{(f \sigma)} \}} \prod^{N_{f \sigma}}_{i=1}
    a^2(\mib{k}^{(f \sigma)}_i)\\
    &=\sum_{\{ \mib{k}^{(f \sigma)} \}}
    \exp
    \left\{-\sum^{N_{f \sigma}}_{i=1}\left[-\ln a^2(\mib{k}^{(f \sigma)}_i)\right]
    \right\}\\
    &=\exp[-F(N_{f \sigma})].
  \end{split}
\end{equation}
$Z(N_{f \sigma})$ is the partition function of the canonical ensemble
for the system with the dispersion
$\varepsilon_{\mib{k}}=-\ln a^2(\mib{k})$
with the constraint $k<k_{\text{F}}$ at temperature 1.
$F(N_{f \sigma})$ is the free energy of this fictitious system.
Then, by using the Stirling formula,
we rewrite the normalization factor as
\begin{equation}
  \langle \psi | \psi \rangle
  \simeq
  \sum_{N_{f \uparrow} N_{f \downarrow} D_{c \uparrow} D_{c \downarrow}}
  \exp[L f(n_{f \uparrow}, n_{f \downarrow}, d_{c \uparrow}, d_{c \downarrow})],
\end{equation}
where $n_{f \sigma}=N_{f \sigma}/L$ and $d_{c \sigma}=D_{c \sigma}/L$.
In the summation, the most important terms should satisfy
\begin{equation}
  \frac{\partial f(n_{f \uparrow}, n_{f \downarrow}, d_{c \uparrow}, d_{c \downarrow})}
  {\partial d_{c \sigma}}=0,
  \label{eq:ddc}
\end{equation}
and
\begin{equation}
  \frac{\partial f(n_{f \uparrow}, n_{f \downarrow}, d_{c \uparrow}, d_{c \downarrow})}
  {\partial n_{f \sigma}}=0.
  \label{eq:dnf}
\end{equation}
From eq.~\eqref{eq:ddc}, we obtain
\begin{equation}
  g^2=\frac{d_{c \sigma}(1-n_f-n_{c \sigma}+d_{c \sigma})}
  {(n_f-d_{c \sigma})(n_{c \sigma}-d_{c \sigma})},
  \label{eq:g}
\end{equation}
where $n_{c \sigma}=n_{\sigma}-n_{f \sigma}$ with $n_{\sigma}=N_{\sigma}/L$.
This is the same form as that in the Hubbard model,~\cite{Gutzwiller1965}
if we regard
$n_{c \sigma}$ as $n^{\text{H}}_{\sigma}$,
$n_f$ as $n^{\text{H}}_{\bar{\sigma}}$,
and $d_{c \sigma}$ as $d^{\text{H}}$,
where $n^{\text{H}}_{\sigma}$ and  $d^{\text{H}}$
are the numbers of $\sigma$-spin electrons and doubly occupied sites
per lattice site, respectively, in the Hubbard model,
and $\bar{\sigma}$ denotes the opposite spin of $\sigma$.
From eq.~\eqref{eq:dnf}, we obtain
\begin{equation}
  e^{\mu(n_{f \sigma})}
  =
  \frac{n^2_f(n_{c \sigma}-d_{c \sigma})(1-n_{c \sigma})
    (1-n_f-n_{c \bar{\sigma}}+d_{c \bar{\sigma}})}
  {(1-n_{f \sigma})(1-n_f)n_{c \sigma}(n_f-d_{c \sigma})(n_f-d_{c \bar{\sigma}})},
\end{equation}
where $\mu$ is the chemical potential for the fictitious system
defined as
\begin{equation}
  \mu(n_{f \sigma})=\frac{d F(N)}{d N}\bigg\rvert_{N=N_{f \sigma}}.
\end{equation}
%
In the following, we assume a paramagnetic state, i.e.,
$n_{f \sigma}=n_f/2$, $n_{c \sigma}=n_c/2=(n-n_f)/2$,
and $d_{c \sigma}=d/2$,
and optimize the wave function so that it has the lowest energy.
In the following,
we regard $d$ as a variational parameter
instead of $g$ by using eq.~\eqref{eq:g}.
From the definition of the chemical potential in the grand canonical ensemble,
the following equation should be satisfied
\begin{equation}
  \begin{split}
    n_f/2&=\frac{1}{L}\sum_{k<k_{\text{F}}}\frac{1}{1+e^{-\ln a^2(\mib{k})-\mu(n_f/2)}}\\
    &=\frac{1}{L}\sum_{k<k_{\text{F}}}\frac{a^2(\mib{k})}{q^{-1}+a^2(\mib{k})},
  \end{split}
  \label{eq:nf1}
\end{equation}
where we have introduced $q=e^{\mu(n_f/2)}$.

If we set $g=1$, that is,
if we ignore the correlation between the conduction and $f$ electrons,
we obtain $d=n_c n_f$ and
$q^{-1}=(1-n_f/2)/(1-n_f)$,
which is the renormalization factor given in eq.~\eqref{eq:Rice-Ueda}.
Our theory is reduced to the previous Gutzwiller method for the PAM
by setting $g=1$.
This can also be checked for other quantities,
such as renormalization factors,
which will be derived in the following.

Next, we evaluate the kinetic energy.
The effect of the annihilation operator on the variational wave function
is written as
\begin{equation}
  \begin{split}
    c_{\mib{r}^{\prime} \uparrow}| \psi \rangle
    =&\sum_{N_{f \uparrow} N_{f \downarrow}}
    \sum_{\{\mib{r}\} \{\mib{k}^{(f)}\}}
    g^D
    (-1)^{\text{perm}(\{\mib{k}^{(f)}\})} \\
    &\times
    \prod^{N_{f \uparrow}}_{i=1} a(\mib{k}^{(f \uparrow)}_i)
    \prod^{N_{f \downarrow}}_{j=1} a(\mib{k}^{(f \downarrow)}_j) \\
    &\times
    \det{}^{(\mib{r}^{\prime})}
    [\varphi_{\mib{k}^{(c \uparrow)}}(\mib{r}^{(c \uparrow)})]
    \det[\varphi_{\mib{k}^{(f \uparrow)}}(\mib{r}^{(f \uparrow)})] \\
    &\times
    \det[\varphi_{\mib{k}^{(c \downarrow)}}(\mib{r}^{(c \downarrow)})]
    \det[\varphi_{\mib{k}^{(f \downarrow)}}(\mib{r}^{(f \downarrow)})] \\
    &\times
    | \{\mib{r}^{(c \uparrow)}_1 \cdots
    \mib{r}^{(c \uparrow)}_{N_{c \uparrow}-1}\}
    \{\mib{r}^{(c \downarrow)}\} \{\mib{r}^{(f)}\} \rangle,
  \end{split}
\end{equation}
where
\begin{equation}
  \begin{split}
    &\det{}^{(\mib{r}^{\prime})}
    [\varphi_{\mib{k}^{(c \uparrow)}}(\mib{r}^{(c \uparrow)})]\\
    =
    &\begin{vmatrix}
      \varphi_{\mib{k}^{(c \uparrow)}_1}(\mib{r}^{\prime}) &
      \varphi_{\mib{k}^{(c \uparrow)}_1}(\mib{r}^{(c \uparrow)}_1) & \\
      \varphi_{\mib{k}^{(c \uparrow)}_2}(\mib{r}^{\prime}) &
      \ddots & \\
      & &
      \varphi_{\mib{k}^{(c \uparrow)}_{N_{c \uparrow}}}
      (\mib{r}^{(c \uparrow)}_{N_{c \uparrow}-1})
    \end{vmatrix}.
  \end{split}
  \label{eq:determ}
\end{equation}
Thus, we need to introduce another approximation to evaluate
the determinant eq.~\eqref{eq:determ}.
By using eq.~\eqref{eq:average2},
we obtain, for $\mib{r} \ne \mib{r}^{\prime}$,
\begin{equation}
  \begin{split}
    \langle \psi |
    c^{\dagger}_{\mib{r} \uparrow}
    c_{\mib{r}^{\prime} \uparrow}
    |\psi \rangle
    \simeq
    &q_{c \uparrow}  
    \sum_{N_{f \uparrow} N_{f \downarrow} D_{c \uparrow} D_{c \downarrow}}
    g^{2D}
    X(N_{f \uparrow},N_{f \downarrow})\\
    &\times
    Y(N_f,N_{c \uparrow},D_{c \uparrow})
    Y(N_f,N_{c \downarrow},D_{c \downarrow})\\
    &\times
    \sum_{\{ \mib{k}^{(f)} \}}
    \prod^{N_{f \uparrow}}_{i=1} a^2(\mib{k}^{(f\uparrow)}_i)
    \prod^{N_{f \downarrow}}_{j=1} a^2(\mib{k}^{(f\downarrow)}_j)\\
    &\times
    \sum^{N_{c \uparrow}}_{l=1}
    \varphi^*_{\mib{k}^{(c \uparrow)}_l}(\mib{r})
    \varphi_{\mib{k}^{(c \uparrow)}_l}(\mib{r}^{\prime}).
  \end{split}
  \label{eq:cc_realspace}
\end{equation}
The renormalization factor $q_{c \sigma}$ is given by
\begin{equation}
  \begin{split}
    q_c=q_{c \sigma}
    =&\frac{1}{n_{c \sigma}(1-n_{c \sigma})}\\
    &\times\biggl[\sqrt{(n_{c \sigma}-d_{c \sigma})
      (1-n_f-n_{c \sigma}+d_{c \sigma})}\\
    &+\sqrt{d_{c \sigma}(n_f-d_{c \sigma})}\biggr]^2.
  \end{split}
\end{equation}
$q_{c \sigma}$ has the same form as
the renormalization factor $q^{\text{H}}_{\sigma}$
in the Hubbard model~\cite{Gutzwiller1965}
as in the case of the Gutzwiller parameter $g$.
From eq.~\eqref{eq:cc_realspace},
we can evaluate the momentum distribution function
$n_c(\mib{k})=n_{c \uparrow}(\mib{k})
=\langle c^{\dagger}_{\mib{k} \uparrow} c_{\mib{k} \uparrow} \rangle
=\langle \psi |c^{\dagger}_{\mib{k} \uparrow} c_{\mib{k} \uparrow} | \psi \rangle
/\langle \psi | \psi \rangle$.
In this evaluation,
we need to calculate 
$\sum_{\{ \mib{k}^{(f) \uparrow} \}}
\prod^{N_{f \uparrow}}_{i=1} a^2(\mib{k}^{(f\uparrow)}_i)$
with the restriction $\mib{k} \notin \{ \mib{k}^{(f) \uparrow} \}$.
We can accomplish it with the aid of eq.~\eqref{eq:prod}.
The result is
\begin{equation}
  n_{c \uparrow}(\mib{k})
  =
  \begin{cases}
    (1-q_{c \uparrow})n_{c \uparrow}+\Delta n_c(\mib{k}) & \text{for $k<k_{\text{F}}$} \\
    (1-q_{c \uparrow})n_{c \uparrow}                 & \text{for $k>k_{\text{F}}$}
  \end{cases},
  \label{eq:nck}
\end{equation}
where
\begin{equation}
  \Delta n_c(\mib{k})=q_c\frac{q^{-1}}{q^{-1}+a^2(\mib{k})}.
\end{equation}
In a similar way, we obtain
$n_f(\mib{k})=n_{f \uparrow}(\mib{k})
=\langle f^{\dagger}_{\mib{k} \uparrow} f_{\mib{k} \uparrow} \rangle$
as
\begin{equation}
  n_{f \uparrow}(\mib{k})
  =
  \begin{cases}
    (1-q_{f \uparrow})n_{f \uparrow}+\Delta n_f(\mib{k}) & \text{for $k<k_{\text{F}}$} \\
    (1-q_{f \uparrow})n_{f \uparrow}                 & \text{for $k>k_{\text{F}}$}
  \end{cases},
  \label{eq:nfk}
\end{equation}
where
\begin{equation}
  \Delta n_f(\mib{k})=q_f\frac{a^2(\mib{k})}{q^{-1}+a^2(\mib{k})}.
\end{equation}
The renormalization factor for an $f$ electron is
\begin{equation}
  q_f=q_{f \sigma}=\frac{1-n_f}{1-n_{f \sigma}}
  q^{(c \uparrow)}_f q^{(c \downarrow)}_f,
\end{equation}
where
\begin{equation}
  \begin{split}
    q^{(c \sigma)}_f
    =&\frac{1}{n_f(1-n_f)}\\
    &\times\Bigl[\sqrt{(n_f-d_{c \sigma})(1-n_f-n_{c \sigma}+d_{c \sigma})}\\
    &+\sqrt{d_{c \sigma}(n_{c \sigma}-d_{c \sigma})}\Bigr]^2.
  \end{split}
\end{equation}
$q^{(c \sigma)}_f$ has the same form as
$q^{\text{H}}_{\sigma}$ in the Hubbard model,~\cite{Gutzwiller1965}
if we regard $n_f$ as $n^{\text{H}}_{\sigma}$,
$n_{c \sigma}$ as $n^{\text{H}}_{\bar{\sigma}}$,
and $d_{c \sigma}$ as $d^{\text{H}}$.
We can also evaluate the mixing term
\begin{equation}
  \langle c^{\dagger}_{i \uparrow} f_{i \uparrow} \rangle
  =q_{cf}\frac{1}{L}\sum_{k<k_{\text{F}}}
  \frac{a(\mib{k})}{q^{-1}+a^2(\mib{k})},
\end{equation}
where the renormalization factor is given by
\begin{equation}
  \begin{split}
    q_{cf}=q_{cf \sigma}&=
    \frac{(n_f-d_{c \sigma})(n_f-d_{c \bar{\sigma}})}{n^2_f(1-n_{c \sigma})}\\
    &\times\left[1+
      \sqrt{\frac{d_{c \bar{\sigma}}(n_{c \bar{\sigma}}-d_{c \bar{\sigma}})}
        {(n_f-d_{c \bar{\sigma}})(1-n_f-n_{c \bar{\sigma}}+d_{c \bar{\sigma}})}}\right].
  \end{split}
\end{equation}

Then, the expectation value of energy
$e=\langle H \rangle/L$ per site is given by
\begin{equation}
  \begin{split}
    e
    =&\frac{2}{L}\sum_{k<k_{\text{F}}}\tilde{\epsilon}_{\mib{k}}\\
    &+\frac{2}{L}\sum_{k<k_{\text{F}}}
    \frac{(\epsilon_f-\tilde{\epsilon}_{\mib{k}})a^2(\mib{k})
      -2\tilde{V}_1a(\mib{k})}{q^{-1}+a^2(\mib{k})}+U_{cf}d,
  \end{split}
  \label{eq:E}
\end{equation}
where
$\tilde{\epsilon}_{\mib{k}}=q_c\epsilon_{\mib{k}}$
and
$\tilde{V}_1=q_{cf}V$.
We minimize the expectation value of energy
with respect to the variational parameters
$a(\mib{k})$ and $d$.
From $\partial e / \partial a(\mib{k})=0$,
we obtain
\begin{equation}
  a(\mib{k})=
  \frac{2\tilde{V}_1}
  {\tilde{\epsilon}_f-\tilde{\epsilon}_{\mib{k}}
    +\sqrt{(\tilde{\epsilon}_f-\tilde{\epsilon}_{\mib{k}})^2+4\tilde{V}^2_2}},
  \label{eq:ak}
\end{equation}
where
$\tilde{V}_2=\sqrt{q}\tilde{V}_1$.
The renormalized $f$-level $\tilde{\epsilon}_f$
should satisfy
\begin{equation}
  \begin{split}
    \epsilon_f-\tilde{\epsilon}_f=
    &-2\tilde{V}^2_2 I_2 q \frac{\partial q^{-1}}{\partial n_f}\\
    &-(I_1-I_4-I_3 \tilde{\epsilon}_f)
    q^{-1}_c\frac{\partial q_c}{\partial n_f}\\
    &+4\tilde{V}^2_2 I_2  q^{-1}_{cf}\frac{\partial q_{cf}}{\partial n_f}.
  \end{split}
  \label{eq:ef}
\end{equation}
The integrals are given by
\begin{equation}
  I_1=\frac{1}{L}\sum_{k<k_{\text{F}}}\tilde{\epsilon}_{\mib{k}},
  \label{eq:I1}
\end{equation}
and
\begin{equation}
  I_{l}=\frac{1}{L}\sum_{k<k_{\text{F}}}
  \frac{(\tilde{\epsilon}_{\mib{k}}-\tilde{\epsilon}_f)^{l-2}}
  {\sqrt{(\tilde{\epsilon}_{\mib{k}}-\tilde{\epsilon}_f)^2+4\tilde{V}^2_2}},
  \label{eq:I}
\end{equation}
for $l=2$--4.
From $\partial e / \partial d=0$,
we obtain
\begin{equation}
  \begin{split}
    U_{cf}=
    &-2\tilde{V}^2_2 I_2 q\frac{\partial q^{-1}}{\partial d}\\
    &-(I_1-I_4-I_3 \tilde{\epsilon}_f) q^{-1}_c\frac{\partial q_c}{\partial d}\\
    &+4\tilde{V}^2_2 I_2  q^{-1}_{cf}\frac{\partial q_{cf}}{\partial d}.
  \end{split}
  \label{eq:Ucf}
\end{equation}
Here, note that
while we take $a(\mib{k})$ and $d$
as independent variables for 
$\partial e / \partial a(\mib{k})=0$
and
$\partial e / \partial d=0$,
we take $n_f$ and $d$
as independent variables for the derivatives
in eqs.~\eqref{eq:ef} and \eqref{eq:Ucf}.
Equation~\eqref{eq:nf1} is rewritten using
eqs.~\eqref{eq:ak} and \eqref{eq:I} as
\begin{equation}
  n_f=\frac{n}{2}+I_3.
  \label{eq:nf2}
\end{equation}
We solve eqs.~\eqref{eq:ef}, \eqref{eq:Ucf}, and \eqref{eq:nf2},
and determine $\tilde{\epsilon}_f$, $d$, and $n_f$.
%
By using eqs.~\eqref{eq:ak}, \eqref{eq:I1}, and \eqref{eq:I},
we can rewrite eq.~\eqref{eq:E} as
\begin{equation}
  e=
  I_1+n_f \epsilon_f+\left(\frac{n}{2}-n_f\right)\tilde{\epsilon}_f
  -I_4-4\tilde{V}^2_2I_2
  +U_{cf}d.
\end{equation}
%
From the above equations,
we find that
the band structure of the conduction band is included only
through the density of states in the present model,
and a physical quantity, such as $n_c(\mib{k})$,
depends on the momentum $\mib{k}$ only through $\epsilon_{\mib{k}}$.

We can evaluate expectation values of physical quantities
in the optimized wave function.
For example,
we obtain the momentum distribution function
$n_c(\mib{k})$ and $n_f(\mib{k})$
using eqs.~\eqref{eq:nck} and \eqref{eq:nfk}, respectively.
An important quantity is the jump
$\Delta n(k_{\text{F}})=\Delta n_c(k_{\text{F}})+\Delta n_f(k_{\text{F}})$
at the Fermi level;
its inverse corresponds to the mass enhancement factor.
In the following,
we call $1/\Delta n(k_{\text{F}})$ the mass enhancement factor.

\begin{figure}
  \includegraphics[width=0.99\linewidth]
  {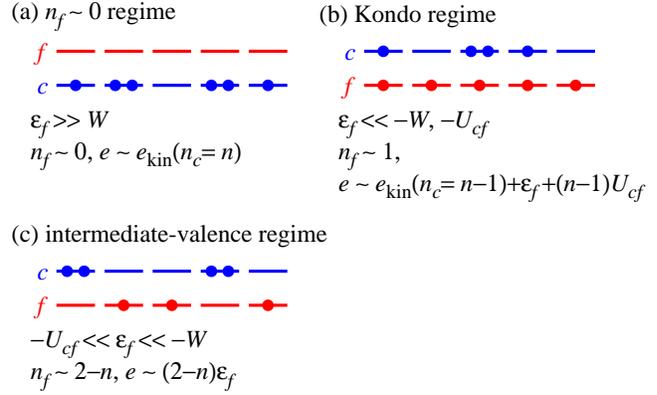}
  \caption{\label{figure:three_regimes}
    (Color online)
    Typical electron configurations in three characteristic regimes:
    (a) $n_f \simeq 0$ regime,
    (b) Kondo regime,
    and
    (c) intermediate-valence regime.
    $e_{\text{kin}}(n_c)$ denotes the kinetic energy per site
    for the free conduction band with $n_c$.
  }
\end{figure}

In the next section, we show the calculated results
for the model with a constant density of states.
Before showing them,
we discuss three characteristic regimes of the model,
schematically presented in Fig.~\ref{figure:three_regimes},
which do not depend on the details of the band structure.
First, we consider a case with $\epsilon_f \gg W$
[Fig.~\ref{figure:three_regimes}(a)],
where $W$ is a typical energy scale of the conduction band
or half of the bandwidth in the next section.
In this case, $n_f \simeq 0$ and the energy $e$ per site
is almost the same as the kinetic energy $e_{\text{kin}}(n_c)$ per site
of the free conduction band with $n_c=n$.
Second, we consider a case with $\epsilon_f \ll -W$, $-U_{cf}$
[Fig.~\ref{figure:three_regimes}(b)].
In this case, $n_f \simeq 1$ and $n_c \simeq n-1$.
The energy is approximately given by
$e \simeq e_{\text{kin}}(n_c=n-1)+\epsilon_f+(n-1)U_{cf}$.
We call this regime the Kondo regime.
For $n_f \rightarrow 1$, we obtain
$q \rightarrow 0$, $q_c \rightarrow 1$, $q_f \rightarrow 0$,
and $a(k_{\text{F}})$ diverges as $a(k_{\text{F}}) \sim q^{-1}$.
By using them, we find that, for $n_f \rightarrow 1$,
$\Delta n_c(k_{\text{F}}) \rightarrow 0$ and 
$\Delta n_f(k_{\text{F}}) \rightarrow 0$,
that is,
the mass enhancement factor becomes large.
This mass enhancement for $n_f \rightarrow 1$
is consistent with the previous result for the PAM.
Third, we consider a case with a moderate $\epsilon_f$
and a large $U_{cf}$, more explicitly,
$-U_{cf} \ll \epsilon_f \ll -W$
[Fig.~\ref{figure:three_regimes}(c)].
In this case, $f$ and conduction electrons tend to avoid each other;
thus,
$n_f+n_c/2 \simeq 1$ and $d \simeq 0$.
That is, $n_f \simeq 2-n$ and $n_c \simeq 2n-2$.
Here, we call this regime the intermediate-valence regime.
In this case, both $f$ and conduction electrons are almost localized,
and the energy is $e \simeq (2-n) \epsilon_f$.
For $n_f+n_c/2 \rightarrow 1$ and $d \rightarrow 0$,
we obtain
$q \rightarrow 0$, $q_c \rightarrow 0$, and $q_f \rightarrow 0$.
By using them,
we can show that the mass enhancement factor becomes large
in this intermediate-valence regime.
This mass enhancement in the intermediate-valence regime
is not realized in the ordinary PAM
and is a result of the effect of $U_{cf}$.

\section{Results}
Now, we show our calculated results.
Here, we consider a simple model
for the kinetic energy:
the density of states per spin is given by
$\rho(\epsilon)=1/(2W)$ for $-W \le \epsilon \le W$;
otherwise, $\rho(\epsilon)=0$.

\begin{figure}
  \includegraphics[width=0.99\linewidth]
  {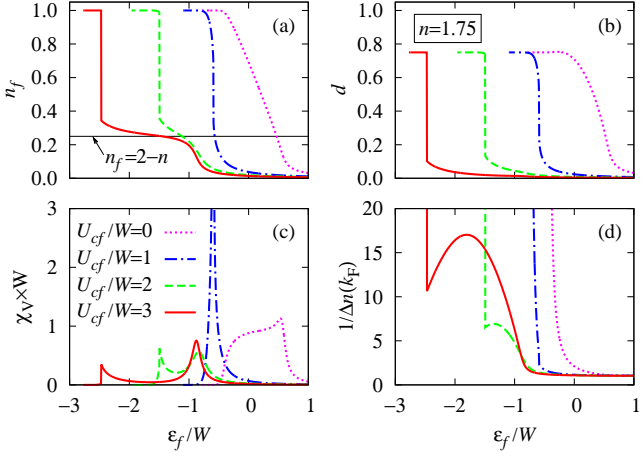}
  \caption{\label{figure:n1.75_V.1}
    (Color online)
    $\epsilon_f$ dependences of
    (a) $n_f$,
    (b) $d$,
    (c) $\chi_{\text{V}}$,
    and
    (d) $1/\Delta n(k_{\text{F}})$
    for $V/W=0.1$ and $n=1.75$.
    $U_{cf}/W=0$ (dotted lines), 1 (dash-dotted lines),
    2 (dashed lines), and 3 (solid lines).
  }
\end{figure}

Figure~\ref{figure:n1.75_V.1}(a) shows
$n_f$ as a function of $\epsilon_f$ for several $U_{cf}$ values
for $V/W=0.1$ and $n=1.75$.
For a large $U_{cf}$, we recognize the three regimes mentioned above.
A first-order phase transition occurs from the Kondo regime
to the intermediate-valence regime
or to the $n_f \simeq 0$ regime
for $U_{cf}/W > 0.89$.
We observe hysteresis by increasing and decreasing $\epsilon_f$
across the first-order phase transition point,
and here we show the values of the state that has the lower energy.
Figure~\ref{figure:n1.75_V.1}(b) shows
the number of interacting electron pairs $d$ through $U_{cf}$ per site.
For a large $U_{cf}$, the conduction and $f$ electrons tend to
avoid each other and $d$ is suppressed.
Figure~\ref{figure:n1.75_V.1}(c) shows
the valence susceptibility $\chi_{\text{V}}=-d n_f/d \epsilon_f$
as a function of $\epsilon_f$.
The valence susceptibility enhances around the boundaries
of three regimes for a large $U_{cf}$.
For $U_{cf}=0$, such a boundary is not so clear.
Figure~\ref{figure:n1.75_V.1}(d) shows
the mass enhancement factor $1/\Delta n(k_{\text{F}})$
as a function of $\epsilon_f$.
In addition to the enhancement for $n_f \rightarrow 1$
as in the ordinary PAM,
we find another region, that is, the intermediate-valence regime
$n_f \simeq 2-n$,
in which the mass enhancement factor becomes large.
This enhancement, in particular, a peak as a function of $\epsilon_f$,
is not expected for the PAM without $U_{cf}$.
Our theory may be relevant to
the large effective mass in the intermediate-valence compounds
$\alpha$-YbAlB$_4$ and $\beta$-YbAlB$_4$
and the nonmonotonic variation in the effective mass under pressure
in CeCu$_2$Si$_2$.

\begin{figure}
  \includegraphics[width=0.99\linewidth]
  {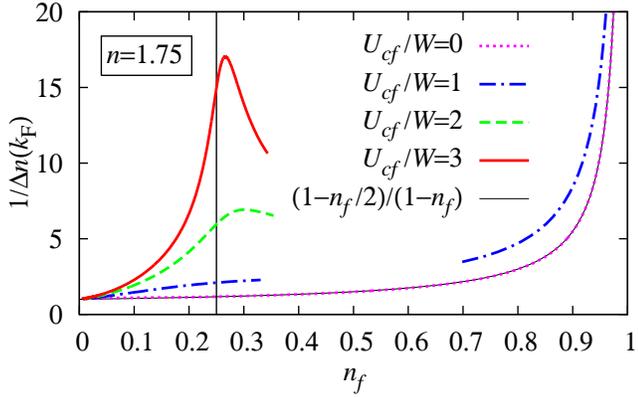}
  \caption{\label{figure:n1.75_V.1_effective_mass_func_of_nf}
    (Color online)
    $1/\Delta n(k_{\text{F}})$ as a function of $n_f$ for $V/W=0.1$ and $n=1.75$.
    $U_{cf}/W=0$ (dotted lines), 1 (dash-dotted lines),
    2 (dashed lines), and 3 (solid lines).
    The thin line is $(1-n_f/2)/(1-n_f)$.
    The vertical line indicates $n_f=2-n$.
  }
\end{figure}

To clearly observe the effect of $U_{cf}$ on the mass enhancement,
we show $1/\Delta n(k_{\text{F}})$ as a function of $n_f$
in Fig.~\ref{figure:n1.75_V.1_effective_mass_func_of_nf}.
The thin line, which almost overlaps with the $U_{cf}=0$ data,
represents the mass enhancement factor, given by eq.~\eqref{eq:Rice-Ueda},
i.e.,
$(1-n_f/2)/(1-n_f)$
obtained for the PAM with $U_{cf}=0$ and $g=1$.
Note that, in the present theory,
$g \ne 1$ even for $U_{cf}=0$.
By increasing $U_{cf}$, $1/\Delta n(k_{\text{F}})$ becomes large,
particularly in the intermediate-valence regime $n_f \simeq 2-n$.

\begin{figure}
  \includegraphics[width=0.99\linewidth]
  {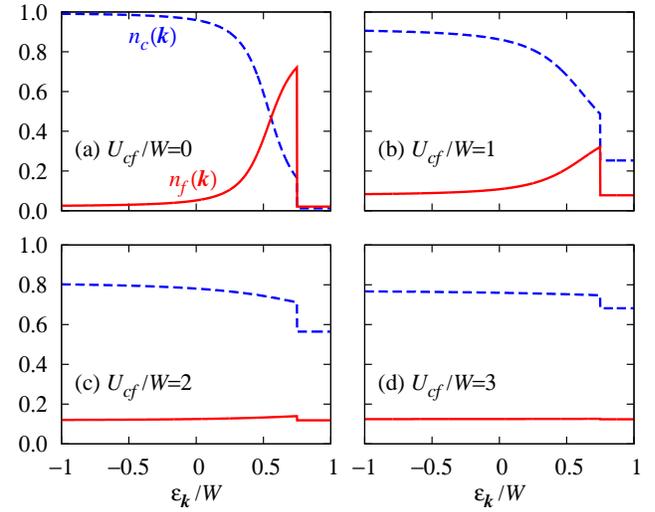}
  \caption{\label{figure:n1.75_V.1_momentum_distribution}
    (Color online)
    Momentum distribution functions
    $n_c(\mib{k})$ (dashed lines)
    and
    $n_f(\mib{k})$ (solid lines)
    as functions of $\epsilon_{\mib{k}}$
    for $V/W=0.1$, $n=1.75$, and $n_f=2-n=0.25$.
    (a) $U_{cf}/W=0$,
    (b) $U_{cf}/W=1$,
    (c) $U_{cf}/W=2$,
    and
    (d) $U_{cf}/W=3$.
  }
\end{figure}

In Fig.~\ref{figure:n1.75_V.1_momentum_distribution}, we show
the momentum distribution functions $n_c(\mib{k})$ and $n_f(\mib{k})$
for $n=1.75$ and $n_f=2-n=0.25$
for several values of $U_{cf}$.
For $U_{cf}=0$, the jump at the Fermi energy $\epsilon_{k_{\text{F}}}=(n-1)W=0.75W$
is much larger for $n_f(k_{\text{F}})$ than for $n_c(k_{\text{F}})$,
that is,
the quasiparticle weight is mainly composed of the $f$-electron contribution.
For a large $U_{cf}$,
the jump becomes small for both $n_f(k_{\text{F}})$ and $n_c(k_{\text{F}})$,
and the mass enhancement factor becomes large,
as shown in Fig.~\ref{figure:n1.75_V.1_effective_mass_func_of_nf}.

\begin{figure}
  \includegraphics[width=0.99\linewidth]
  {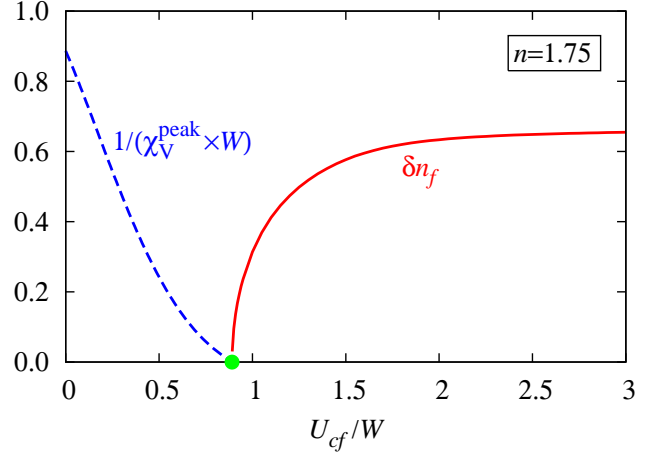}
  \caption{\label{figure:n1.75_V.1_critical_point}
    (Color online)
    Inverse of peak value $\chi^{\text{peak}}_{\text{V}}$ (dashed line)
    of valence susceptibility [see Fig.~\ref{figure:n1.75_V.1}(c)]
    and
    jump $\delta n_f$ (solid line)
    in $n_f$ [see Fig.~\ref{figure:n1.75_V.1}(a)]
    at first-order phase transition
    as functions of $U_{cf}$
    for $V/W=0.1$ and $n=1.75$.
    The circle represents the critical point.
  }
\end{figure}

Figure~\ref{figure:n1.75_V.1_critical_point} shows
how we determine the critical point of the valence transition.
In this figure, we draw the inverse of
the peak $\chi^{\text{peak}}_{\text{V}}$ of the valence susceptibility
and the jump $\delta n_f$ in $n_f$ at the first-order valence transition.
Both of them should become zero at the critical point,
and indeed,
we find that they become zero at the same $U_{cf}$.

\begin{figure}
  \includegraphics[width=0.99\linewidth]
  {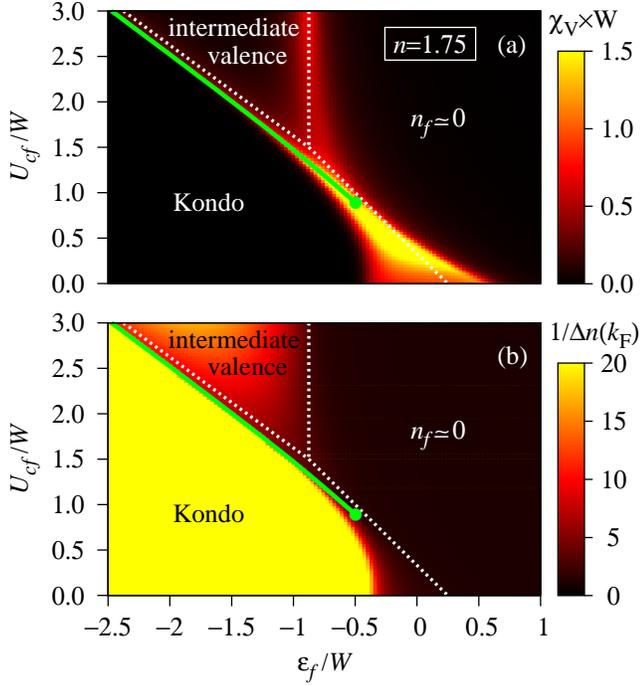}
  \caption{\label{figure:n1.75_V.1_chiV_effective_mass}
    (Color online)
    (a) $\chi_{\text{V}}$
    and
    (b) $1/\Delta n(k_{\text{F}})$ 
    as functions of $\epsilon_f$ and $U_{cf}$
    for $n=1.75$ with $V/W=0.1$.
    The solid lines represent the first-order
    valence transition line.
    The solid circles denote the critical point of
    the valence transition.
    The dotted lines indicate crossover lines
    determined by comparing the energies of the three extreme states (see text).
  }
\end{figure}

In Fig.~\ref{figure:n1.75_V.1_chiV_effective_mass}(a),
we show the valence susceptibility $\chi_{\text{V}}$
as a function of $\epsilon_f$ and $U_{cf}$ for $n=1.75$.
In this figure,
we also draw the first-order valence transition line
and its critical point.
The crossover lines, represented by the dotted lines,
are determined by comparing the energies of the three extreme states:
$n_f=0$, $n_f=1$, and $n_f+n_c/2=1$ with $d=0$.
The crossover lines are given by
\begin{equation}
  \epsilon_f=-(n-1)U_{cf}+e_{\text{kin}}(n_c=n)-e_{\text{kin}}(n_c=n-1),
  \label{eq:Kondo-nf0}
\end{equation}
between the Kondo and $n_f \simeq 0$ regimes, by
\begin{equation}
  \epsilon_f=\frac{e_{\text{kin}}(n_c=n)}{2-n},
\end{equation}
between the intermediate-valence and $n_f \simeq 0$ regimes,
and by
\begin{equation}
  \epsilon_f=-U_{cf}-\frac{e_{\text{kin}}(n_c=n-1)}{n-1},
\end{equation}
between the Kondo and intermediate-valence regimes.
The crossover line
between the intermediate-valence and $n_f \simeq 0$ regimes
does not depend on $U_{cf}$.
The other crossover lines are straight lines with finite slopes.
Between the Kondo and $n_f \simeq 0$ regimes,
the slope is $-1/(n-1)$
and does not depend on the band structure.
Between the Kondo and intermediate-valence regimes,
the slope is $-1$ independent of both the band structure
and filling $n$.~\cite{Watanabe2006}
The region where $\chi_{\text{V}}$ becomes large
is captured well by the crossover lines
obtained by such a simple consideration.
The first-order valence transition occurs
only from the Kondo to intermediate-valence or to $n_f \simeq 0$ regimes
within the $U_{cf}$ range presented here.
Note that
the valence transition can occur
also between the intermediate-valence regime and the $n_f \simeq 0$ regime
for a smaller $n$.~\cite{Kubo2011}
Figure~\ref{figure:n1.75_V.1_chiV_effective_mass}(b) shows
the mass enhancement factor $1/\Delta n(k_{\text{F}})$
as a function of $\epsilon_f$ and $U_{cf}$.
A large mass enhancement occurs in the intermediate-valence regime
in addition to the Kondo regime.
Here, note that
the large mass enhancement occurs in the middle of
the intermediate-valence regime.
Thus, this enhancement is not due to valence fluctuations.
In CeCu$_2$Si$_2$,
the effective mass has a peak before
the superconducting transition temperature becomes maximum under pressure,
and which is consistent with our theory
provided that the pairing interaction of superconductivity
is mediated by the valence fluctuations.
The situation will also be similar for CeCu$_2$Ge$_2$
if we can subtract the contributions of magnetic fluctuations.

\begin{figure}
  \includegraphics[width=0.99\linewidth]
  {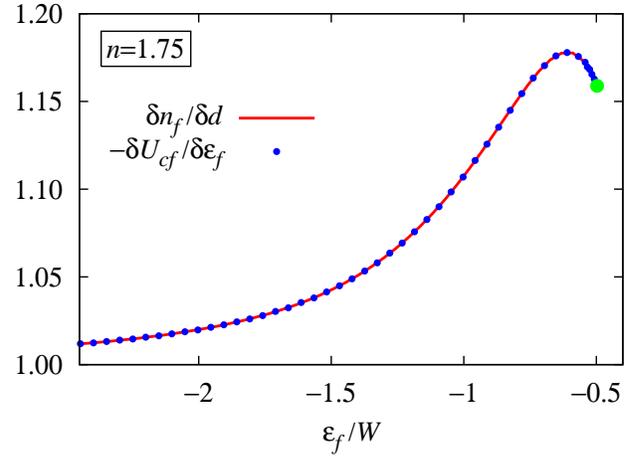}
  \caption{\label{figure:n1.75_V.1_ClausiusClapeyron}
    (Color online)
    Ratio $\delta n_f / \delta d$
    (solid line)
    of jumps [see Figs.~\ref{figure:n1.75_V.1}(a) and \ref{figure:n1.75_V.1}(b)]
    at the first-order phase transition
    and
    slope $-\delta U_{cf} / \delta \epsilon_f$
    (small circles)
    of first-order phase transition line
    (see Fig.~\ref{figure:n1.75_V.1_chiV_effective_mass})
    for $V/W=0.1$ and $n=1.75$.
    The large circle indicates the critical point.
  }
\end{figure}

Finally, to verify the consistency of the present theory,
we check the Claudius-Clapeyron relation
for the first-order valence transition.~\cite{Watanabe2004,Watanabe2006}
This relation is given by
\begin{equation}
  \frac{\delta n_f}{\delta d}
  =-\frac{\delta U_{cf}}{\delta \epsilon_f},
  \label{eq:ClausiusClapeyron}
\end{equation}
where $\delta d$ denotes the jump in $d$ at the valence transition,
and $\delta U_{cf}/\delta \epsilon_f$
is the slope of the valence transition line.
In Fig.~\ref{figure:n1.75_V.1_ClausiusClapeyron},
we show the values of the quantities
on the left and right sides of eq.~\eqref{eq:ClausiusClapeyron}.
We can clearly see that
the Clausius-Clapeyron relation holds
in the present theory.
Note that the Clausius-Clapeyron relation also holds
for the crossover lines mentioned above.
For example,
$n_f \simeq 1$ and $d \simeq n-1$ for the Kondo regime
and $n_f \simeq 0$ and $d \simeq 0$ for the $n_f \simeq 0$ regime,
and then, between these two regimes,
$\delta n_f / \delta d \simeq 1/(n-1)$.
It is the slope $-\delta U_{cf}/\delta \epsilon_f$
for that crossover line given by eq.~\eqref{eq:Kondo-nf0}.

\section{Summary and Discussion}
We have studied the extended periodic Anderson model
with $U_{cf}$ by Gutzwiller approximation.
We have found that the three regimes, that is,
the Kondo, intermediate-valence, and $n_f \simeq 0$ regimes,
are clearly defined for a large $U_{cf}$.
Then, we have found that, in the intermediate-valence regime,
the effective mass is enhanced substantially.
According to the present theory,
the large mass enhancement in the intermediate-valence regime
indicates a large $U_{cf}$.
Thus, our theory provides helpful information
for searching a superconductor
with valence-fluctuation-mediated pairing.

In this study, we have not considered
the possible instability toward a spin-density-wave state
and a charge-density-wave state.
Such a state would be realized
in a portion of the parameter space, particularly,
in a lattice without geometric frustration.~\cite{Sugibayashi2008,Yoshida2011}
The extension of the present theory to such states is a future problem.
In the present theory for a uniform state,
the effect of a lattice structure is included
only through the density of states of the conduction band.
Thus, our results may change little for a frustrated lattice
with a similar density of states
even if we consider the possibility of the density-wave states.

In our theory,
we expand both the conduction- and $f$-electron states
in the basis states in real space;
thus,
it will be possible to include
the onsite correlation between conduction electrons
and other short-range correlations.
These extensions are future problems.

\section*{Acknowledgment}
This work is supported by
a Grant-in-Aid for Young Scientists (B) from
the Japan Society for the Promotion of Science.

\appendix

\section{Approximation for Determinants \label{sec:determ}}
In this appendix, we introduce approximations
for determinants to evaluate expectation values
in the variational wave function.
Although most of them have been derived in ref.~\citen{Fazekas1987},
we repeat them for the readers' convenience.

We consider the state
\begin{equation}
  \lvert \mib{k}_1 \cdots \mib{k}_N \rangle
  =c^{\dagger}_{\mib{k}_1} \cdots c^{\dagger}_{\mib{k}_N}
  \lvert 0 \rangle,
\end{equation}
where $c^{\dagger}_{\mib{k}_i}$ denotes the creation operator
of a spinless fermion with the momentum $\mib{k}_i$.
From eq.~\eqref{eq:basis_state},
we obtain
\begin{equation}
  \begin{split}
    &\langle \mib{k}^{\prime}_1 \cdots \mib{k}^{\prime}_N
    | \mib{k}_1 \cdots \mib{k}_N \rangle\\
    =&\sum_{\{ \mib{r} \}}
    \det[\varphi^*_{\mib{k}^{\prime}}(\mib{r})]
    \det[\varphi_{\mib{k}}(\mib{r})]\\
    =&\delta_{\mib{k}_1 \mib{k}^{\prime}_1}  \cdots
    \delta_{\mib{k}_N \mib{k}^{\prime}_N}.
  \end{split}
\end{equation}
Then, we approximate each product of determinants
by the average, that is,
\begin{equation}
  \lvert\det[\varphi_{\mib{k}}(\mib{r})]\rvert^2
  \simeq \frac{1}{_LC_N},
  \label{eq:average}
\end{equation}
and
\begin{equation}
  \det[\varphi^*_{\mib{k}^{\prime}}(\mib{r})]
  \det[\varphi_{\mib{k}}(\mib{r})]
  \simeq 0,
  \label{eq:orthogonal}
\end{equation}
for $\{ \mib{k} \} \ne \{ \mib{k}^{\prime} \}$.

For the kinetic energy, we need to evaluate another type of determinant.
We consider
\begin{equation}
  \begin{split}
    c_{\mib{r}^{\prime}}| \mib{k}_1 \cdots \mib{k}_N \rangle
    =\sum_{\{ \mib{r} \}}
    &\det[\varphi_{\mib{k}}(\mib{r})]c_{\mib{r}^{\prime}}
    | \mib{r}_1 \cdots \mib{r}_N \rangle\\
    =\sum_{\{ \mib{r} \} \not\ni \mib{r}^{\prime}}
    &\det{}^{(\mib{r}^{\prime})}[\varphi_{\mib{k}}(\mib{r})]
    | \mib{r}_1 \cdots \mib{r}_{N-1} \rangle,
  \end{split}
\end{equation}
where
\begin{equation}
    \det{}^{(\mib{r}^{\prime})}[\varphi_{\mib{k}}(\mib{r})]
    =
    \begin{vmatrix}
      \varphi_{\mib{k}_1}(\mib{r}^{\prime}) &
      \varphi_{\mib{k}_1}(\mib{r}_1) & \\
      \varphi_{\mib{k}_2}(\mib{r}^{\prime}) &
      \ddots & \\
      & &
      \varphi_{\mib{k}_N}(\mib{r}_{N-1})
    \end{vmatrix}.
\end{equation}
Then, for $\mib{r}^{\prime} \ne \mib{r}^{\prime \prime}$,
\begin{equation}
  \begin{split}
    &\langle \mib{k}_1 \cdots \mib{k}_N |
    c^{\dagger}_{\mib{r}^{\prime}} c_{\mib{r}^{\prime \prime}}
    | \mib{k}_1 \cdots \mib{k}_N \rangle\\
    =&\sum_{\{ \mib{r} \} \not\ni \mib{r}^{\prime}, \mib{r}^{\prime \prime}}
    \det{}^{(\mib{r}^{\prime})}[\varphi^*_{\mib{k}}(\mib{r})]
    \det{}^{(\mib{r}^{\prime \prime})}[\varphi_{\mib{k}}(\mib{r})].
  \end{split}
  \label{eq:detrp_detrpp}
\end{equation}
On the other hand, by using the expansion
\begin{equation}
  c_{\mib{r}^{\prime}}=\sum_{\mib{k}^{\prime}}
  \varphi_{\mib{k}^{\prime}}(\mib{r}^{\prime})c_{\mib{k}^{\prime}},
\end{equation}
we obtain
\begin{equation}
  \begin{split}
    &\langle \mib{k}_1 \cdots \mib{k}_N |
    c^{\dagger}_{\mib{r}^{\prime}} c_{\mib{r}^{\prime \prime}}
    | \mib{k}_1 \cdots \mib{k}_N \rangle\\
    =&\sum_{\mib{k}^{\prime}}
    \varphi^*_{\mib{k}^{\prime}}(\mib{r}^{\prime})
    \varphi_{\mib{k}^{\prime}}(\mib{r}^{\prime \prime})
    \langle \mib{k}_1 \cdots \mib{k}_N |
    c^{\dagger}_{\mib{k}^{\prime}} c_{\mib{k}^{\prime}}
    | \mib{k}_1 \cdots \mib{k}_N \rangle\\
    =&\sum^N_{i=1}
    \varphi^*_{\mib{k}_i}(\mib{r}^{\prime})
    \varphi_{\mib{k}_i}(\mib{r}^{\prime \prime}).
  \end{split}
\end{equation}
Then, we approximate the products of determinants in eq.~\eqref{eq:detrp_detrpp}
by their average:
\begin{equation}
  \begin{split}
    &\det{}^{(\mib{r}^{\prime})}[\varphi^*_{\mib{k}}(\mib{r})]
    \det{}^{(\mib{r}^{\prime \prime})}[\varphi_{\mib{k}}(\mib{r})]\\
    \simeq&
    \frac{1}{_{L-2}C_{N-1}}
    \sum^N_{i=1}
    \varphi^*_{\mib{k}_i}(\mib{r}^{\prime})
    \varphi_{\mib{k}_i}(\mib{r}^{\prime \prime}).
  \end{split}
  \label{eq:average2}
\end{equation}

\section{Evaluation of $\sum\prod a^2(\mib{k})$ with Restriction}
In the canonical ensemble for an $N$ free-electron system
with dispersion $\varepsilon_{\mib{k}}$ at temperature $1/\beta$,
the hole distribution function is given by
\begin{equation}
  \langle c_{\mib{k}} c^{\dagger}_{\mib{k}} \rangle
  =\sum_{\{\mib{k}^{\prime}\} \not\ni \mib{k}}
  e^{-\beta\sum^N_{i=1} \varepsilon_{\mib{k}^{\prime}_i}}/Z(N),
\end{equation}
where $Z(N)$ is the partition function.
It should be equivalent to that in the grand canonical ensemble,
\begin{equation}
  \frac{e^{\beta (\varepsilon_{\mib{k}}-\mu)}}
  {1+e^{\beta (\varepsilon_{\mib{k}}-\mu)}},
\end{equation}
where $\mu$ is the chemical potential,
and thus,
\begin{equation}
  \sum_{\{\mib{k}^{\prime}\} \not\ni \mib{k}}
  e^{-\beta\sum^N_{i=1} \varepsilon_{\mib{k}^{\prime}_i}}
  =
  \frac{e^{-\beta \mu}}{e^{-\beta \mu}+e^{-\beta \varepsilon_{\mib{k}}}}
  Z(N).
\end{equation}
By putting
$\varepsilon_{\mib{k}}=-\ln a^2(\mib{k})$, $\beta=1$, and $e^{\mu}=q$,
we obtain
\begin{equation}
  \sum_{\{\mib{k}^{\prime}\} \not\ni \mib{k}}
  \prod^N_{i=1}a^2(\mib{k}^{\prime}_i)
  =\frac{q^{-1}}{q^{-1}+a^2(\mib{k})}
  Z(N).
  \label{eq:prod}
\end{equation}


\end{document}